\documentstyle[preprint,aps,prl]{revtex}
\begin{document}
\draft
\preprint{To appear in Phys.~Rev.~Lett.}
\title{
Shape-independent scaling of excitonic confinement \\
in realistic quantum wires
}
\author{Fausto Rossi, Guido Goldoni, and Elisa Molinari}
\address{
Istituto Nazionale Fisica della Materia (INFM), and
Dipartimento di Fisica, Universit\`a di Modena, I-41100 Modena, Italy}
\date{\today}
\maketitle
\begin{abstract}

The scaling of exciton binding energy in semiconductor quantum wires is
investigated theoretically through a non-variational, fully
three-dimensional approach for a wide set of realistic state-of-the-art
structures. We find that in the strong confinement limit
the same potential-to-kinetic energy ratio holds
for quite different wire cross-sections and compositions. As a consequence,
a universal (shape- and composition-independent) parameter can be
identified that governs the scaling of the binding energy with size.
Previous indications that the shape of the wire
cross-section may have important effects on exciton binding are discussed
in the light of the present results.

\end{abstract}
\clearpage
\narrowtext
The achievement of large exciton binding energies ($E_b$) is considered an
important goal in the field of semiconductor nanostructures. On the one
hand, large values of $E_b$ are the result of the enhanced Coulomb coupling
between electrons and holes due to their localization in the nanostructure
and, therefore, they provide a clear fingerprint of low-dimensional
confinement. On the other hand, large exciton binding energies are a
prerequisite for exploiting excitonic nonlinearities in optical devices
(e.g. switches and modulators) that can operate efficiently at room
temperature.

For ideal two-dimensional (2D) systems, the binding energy of the
ground-state exciton is four times the three-dimensional (3D) effective
Rydberg. In quantum wells (QWs), $E_b$ has been indeed observed to approach
this limit when the well thickness is progressively reduced \cite{2D}.
In the ideal one-dimensional (1D) limit $E_b$ diverges \cite{elliot-loudon},
suggesting that exciton binding energies of quasi-1D (q1D) systems can be
in principle increased by extreme quantum confinement much beyond the 2D
limit \cite{ogawa}.
Moreover, while for the Coulomb interaction
the virial theorem limits $E_b$ to just half of the potential (Coulomb)
energy, the same limitation does not hold in the presence of an additional
confining potential. In other words, if the only interactions are due
to Coulomb forces, the virial theorem implies that the modulus of the ratio
between average potential and kinetic energies is  $\alpha=2$; however,
deviations from this condition are possible in q1D structures, and one might
hope to obtain a more convenient
ratio (i.e. larger $\alpha$) by proper geometrical and
compositional tailoring, thereby enhancing $E_b$.

Large exciton binding energies, clearly indicating additional confinement
with respect to QWs of comparable confinement length $L$, have been indeed
observed
in artificial semiconductor wires fabricated by different
techniques
\cite{V2,T3,V1,T1,T2,V3}.
However, the dependence of $E_b$
on the shape and height of the confining potential, and its scaling with
size are still highly controversial. Based on theoretical predictions
obtained through variational calculations
for model wire geometries, the scaling of $E_b$ is expected to be governed
by the
extension of the single-particle wavefunctions in the plane perpendicular
to the free wire direction (which in most cases simply reflects the cross
sectional area of the wire for a given barrier height), and to be much less
sensitive to the shape of the wire \cite{Degani-Lefebvre}. On the
experimental side, optical spectra have now been obtained for wires of
different geometries. In particular, direct epitaxial overgrowth techniques
on the cleaved edge of multi-QW samples \cite{T-growth} or on patterned
substrates \cite{V-growth} have recently produced wires of good optical
quality, with T-shaped and V-shaped cross-sections, respectively. Among the
recent experimental papers on such samples, some \cite{V2,T3,V1,V3}
essentially confirm the expected trends for the size dependence of $E_b$, 
while in others ---T-shaped wires \cite{T1,T2}--- the values extracted for 
$E_b$ are apparently much larger than expected on the basis of variational 
calculations \cite{T-variat,Reinecke}.

The open question is whether for a specific shape of the wire cross section
electron-hole Coulomb correlation is actually enhanced, due to effects that
have been neglected in previous theoretical approaches. In this case,
the parameters governing the scaling of exciton binding (if any) would have
to be reconsidered. On the other hand, if this is not the case, one would
still have to explain the inconsistencies between the values of $E_b$
extracted from experimental data on different types of samples.

To address this problem, we make use of a theoretical approach recently
proposed and used to study nonlinear optical spectra of quantum wires
\cite{noi}. The scheme is based on a generalization of the
well-known semiconductor Bloch equations (SBE) to the case of a
multisubband wire structure. In this letter, we focus on the
quasi-equilibrium regime where the solution of the SBE simply reduces to
the solution of the polarization equation. This is performed by direct
numerical evaluation of the polarization eigenvalues and eigenvectors,
which fully determine the absorption spectrum. The main ingredients
entering the calculation are the single-particle energies and
wavefunctions, obtained numerically for an arbitrary 2D confinement
potential which, e.g, can be deduced from TEM images of real samples, as in
Ref. \cite{V1}. Since the proposed approach is based on a full 3D
multisubband description of the electron-hole Coulomb 
interaction,\cite{multisubband} it
allows a direct evaluation of the 3D exciton wave function, thereby
eliminating any assumption on the form of the variational excitonic ground
state, which would hamper the determination of possible shape effects.

The above approach has been applied to realistic V- and T-shaped
quantum wire (V-wire and T-wire) structures. For both geometries, two
different sets of conduction and valence band offsets ($V_e$ and
$V_h$, respectively) have been considered, in order to simulate both
low-$x$ Al$_x$Ga$_{1-x}$As and pure AlAs barrier
compositions \cite{barriere}. In total, we
consider four sets of samples, which we label V1, V2, T1, T2, where V
(T) refers to the wire shape, and 1 (2) refers to the low (high)
barriers. For V-wires, we start from the reference-sample TEM profile
of Ref.~\onlinecite{V1}, and magnify or reduce both confinement
directions by the same scale factor. Each sample is characterized by
the confinement length $L_V$ at the bottom of the V-shaped region. For
T-wires, we consider a set of samples with different values of the
parent QW width $L_T$, which includes the samples of
Ref.~\onlinecite{T2} (here we only show data for T-wires with parent
QWs of equal width). The wire geometries are sketched as insets in
Fig.~1, and Table 1 summarizes the parameters characterizing the four
quantum-wire sets reported in this letter.

Figure 1 shows $E_b$ and the corresponding mean potential energy $\langle
V\rangle$ as a function of the characteristic size parameter of the wire,
$L_V$ or $L_T$. (Here and throughout the paper, the symbol $\langle \ldots
\rangle$ denotes the expectation value over the exciton ground state.) As
expected, both binding and potential energies increase with decreasing
$L_V$ or $L_T$; for samples V2 and T2, corresponding to AlAs barriers, the
excitonic binding is larger compared to the case of low-barrier
samples V1 and T1.

Two important features result from Fig. 1. First, a given value
of $E_b$ corresponds to rather different values of $L_V$ and $L_T$
(note the different scale). This tells us that such size parameters
are not adequate to characterize the actual exciton confinement.
To introduce a more appropriate quantity, we define an effective
exciton Bohr radius
\begin{equation}
a = \left\langle {1\over {\bf r}}\right\rangle^{-1},
\end{equation}
whose inverse is clearly proportional to the potential energy and, for a 3D
bulk semiconductor, coincides with the usual exciton Bohr radius
$a_\circ$. The insets in Fig.~1 show $a$ as a function of the relevant
geometrical parameter, $L_V$ or $L_T$.
A same value of $a$ corresponds to different values of $L_V$ and $L_T$,
with $L_V$ always larger than $L_T$. Note that samples with similar
binding energies correspond to similar values of $a$ (see, e.g., the circled
points, to be discussed below).
The second feature resulting from Fig.~1 is that the ratio of binding to
potential energy is rather constant (shape and barrier independent), and
relatively close to one. This tells us that for all the samples considered
the mean kinetic energy $\langle K\rangle$ is much smaller
(about four times) than the potential energy.

Both features indicate a shape-independent scaling of the exciton
binding energy. Indeed, by plotting the binding energy $E_b$ of all
samples vs the corresponding exciton radius $a$ (Fig.~2), what we
obtain is a universal (shape and barrier independent) curve, $E_b \sim
{1\over a}$. A universal scaling of the mean potential and
kinetic energy~\cite{other-geometries} is apparent in the $\langle
V\rangle$ vs $\langle K\rangle$ plot reported in the inset of Fig.~2;
to a very good approximation, all sets of points for V- and T-wires
fall on a straight line with slope $\alpha$ very close to $4$. For
comparison, we have performed analogous calculations for a set of QWs
(the parameters are defined in Table 1). As shown in Fig.~2, we find
that $E_b$ scales with $a$ similarly to q1D structures, although with
a different prefactor. If $\langle V\rangle$ is plotted vs $\langle
K\rangle$ (inset of Fig.~2), in fact, the points for QWs again fall on
a straight line, but now the slope is $\alpha=2$ within numerical
accuracy.

We can therefore conclude that, for q1D structures in the
strong-confinement regime considered here, the potential-to-kinetic energy
ratio is still a constant. However, its value is found to be twice the value
imposed by the conventional virial theorem in 3D and ideal 2D systems, which
we find to be also followed by QWs of comparable confinement lengths.
In this respect, our findings confirm that q1D confinement is indeed
advantageous for the purpose of obtaining enhanced exciton binding,
and provide a general and quantitative prescription for tailoring $E_b$
by tuning the effective exciton Bohr radius $a$ through the geometrical
size parameters.

At the same time, however, the universal scaling law of Fig.~2
sets a clear limit for the possible effects of choosing different
shapes of the wire cross-section, as long as they correspond to similar
values of the effective Bohr radius $a$. For a given value of $a$,
there is no hope to further increase $E_b$ by tailoring the
potential-to-kinetic energy ratio $\alpha$ through the geometry of the
confining profile.

This last conclusion is in apparent contradiction with some findings
that have been reported recently, based on optical experiments
on different wires. In particular, a very large enhancement of
the exciton binding energy was recently reported in high-quality T-wires,
with estimated $E_b$ reaching values 6-7 times larger than the corresponding
3D effective Rydberg~\cite{T2}. These values are much larger than
our theoretical findings for the same nominal potential 
profile.\cite{meaculpa}
More importantly, the effective Bohr radius for such T-wire geometry
is found to be very close to the value of $a$ obtained for samples of
different shape (V-wires, Ref.~\onlinecite{V1}), where much smaller values
of $E_b$ were reported~\cite{foot2}. Our calculated $E_b$ for such
V- and T-wire samples of comparable $a$ (highlighted by circled points
in Fig.~1 and boldface characters in Table 1) are of course very similar.
We think that the origin of this apparent discrepancy is in the procedure
adopted in Ref.~\onlinecite{T2} to extract $E_b$ from the experimental data.
There, the measured quantity is the energy shift of the T-wire exciton
with respect to the exciton of the parent QW, while $E_b$ is derived
by subtracting quantities that are estimated on the basis of
simplified models. Among these, the largest
approximation is made in the estimate of the energy shift between
the lowest single-particle transitions of the T-wire and the parent
QW: If we perform accurate calculations,
including realistic masses and valence band mixing as described in
Ref.\onlinecite{Goldoni96},
we find that the values used in Ref.~\onlinecite{T2} are largely
underestimated \cite{foot4}, leading to an overestimated value of $E_b$.
Further approximations pointed out by the authors
(e.g. the indirect determination of $L_T$) may also contribute to an
overestimation of $E_b$, but are probably less important. When these
corrections are taken into account, the experimental data of Someya et al.
\cite{T2} are indeed compatible with the present theoretical picture, as
well as with the previous experimental results \cite{V1}.

In summary, we have shown that in strongly confined quantum wires
the average Coulomb and kinetic energies are proportional;
their constant ratio is very close to 4, i.e.,
twice the conventional 'virial' value ---which holds in
homogeneous systems and is found here to apply also to QWs---,
thus allowing enhanced $E_b$.
As a consequence of this same proportionality, the scaling of
$E_b$ is found to be governed by a universal parameter that limits the
possible differences due to variations in the shape of the wire
cross-section.
Our results for realistic V- and T-wires in the
strong confinement regime are consistent with the available
experimental data and offer a guideline for tailoring binding energies in
these structures.

We are grateful to T. Someya for useful discussions, and to
T.L. Reinecke for useful correspondence and for communicating 
results prior to publication. This work was supported in part by the EC 
through the TMR-Network 'Ultrafast'.

%
%
\begin{figure}
\caption{Exciton binding energy $E_b$ and mean potential energy
$\langle V\rangle$ of V-wires (left)
and T-wires (right) for the samples of Table \protect\ref{tab:samples}.
Full dots indicate high barrier samples, empty dots indicate low
barrier samples, according to the legends.
Full lines ($E_b$) and dotted lines ($\langle V\rangle$) are
just a guide to the eye. In the left insets we sketch the
wire geometry, with indication of the relevant geometrical parameter.
In the right insets we show the calculated effective exciton Bohr radius,
$a$, vs the relevant geometrical
parameter. The circled points refer to sample parameters
corresponding to Ref.~\protect\cite{V1} (V-wires) and
Ref.~\protect\cite{T2} (T-wires).
For sample parameters see also Table  \protect\ref{tab:samples}.}
\end{figure}
\begin{figure}
\caption{Exciton binding energy, $E_b$, vs effective exciton Bohr radius,
$a$, for the four sets of V- and T-wires, and for the set of QWs
of Table \protect\ref{tab:samples}. Dashed curves are a fitting to
$1/a$ form.
The inset reports the average potential vs kinetic energy,
falling on a straight line with slope $\alpha\simeq 4$
for all wire samples. Results for QW structures
are also shown for comparison; in this case $\alpha\simeq 2$. Solid lines
are a linear fit to the calculated points.}
\end{figure}
%
%
\begin{table}
\caption{Sample parameters and calculated $E_b$
for the four sets of wires. $V_e$, $V_h$, and $E_b$ are given in meV; $L_V, L_T$
are given in nm. Other parameters are the electron effective mass
$m_e=0.067 m_0$, and the hole effective mass $m_h=0.38 m_0$ along the
[001] crystallographic direction, and $m_h=0.69  m_0$ along the
[110] crystallographic direction, where $m_0$ is the free electron mass.
The values in boldface refer to samples
for which $E_b$ has been experimentally evaluated.}
\label{tab:samples}
\begin{tabular}{c|cr|crrrrrr}
V1 & $V_e$ & 150 & $L_V$ & 3.48  & 5.22  & 6.96  & {\bf 8.70}$^a$  & 10.44 & 12.18 \\
   & $V_h$ & 50  & $E_b$ & 14.12 & 13.91 & 12.81 & {\bf 11.66} & 10.63 & 9.76 \\
\hline
V2 & $V_e$ & 1036 & $L_V$ & 3.48  & 5.22  & 6.96  & 8.70  & 10.44 & 12.18 \\
   & $V_h$ & 558 & $E_b$ & 23.37 & 19.10 & 16.13 & 13.97 & 12.36 & 11.12 \\
\hline
T1 & $V_e$ & 243 & $L_T$ & 3.24  & 4.32  & {\bf 5.40}$^b$  & 6.48  & 7.56  &  \\
   & $V_h$ & 131 & $E_b$ & 14.85 & 13.39 & {\bf 11.63} & 10.63 & 9.82  &  \\
\hline
T2 & $V_e$ & 1036 & $L_T$ & 3.24  & 4.32  & {\bf 5.40}$^b$  & 6.48  & 7.56  &  \\
   & $V_h$ & 558 & $E_b$ & 19.90 & 16.23 & {\bf 13.90} & 12.41 & 11.26 &  \\
\hline
QW & $V_e$ & 1036 & $L$   & 3.24  & 5.40  & 8.64  & 16.20 &       &  \\
   & $V_h$ & 558 & $E_b$ & 10.76 & 9.83  & 8.89  & 7.41  &       &
\end{tabular}
$^a$  parameters corresponding to the sample of Ref.~\protect\cite{V1}.
\newline
$^b$  parameters corresponding to samples of Ref.~\protect\cite{T2}.
\end{table}


\begin{references}
\bibitem{multisubband}
Note that the present multisubband approach includes on the same foot both 
confined and continuum states; in particular, this is essential in 
order to properly describe T-wires, where the 1D ground state is 
relatively close to the 2D continuum ensuing from the parent QWs.
\bibitem{barriere} For the V-wires of Ref.~\protect\onlinecite{V1} the barriers
are constituted by a short period AlAs/GaAs
superlattice. This induces a rather shallow confinement which,
from the point of view of the barrier height, is
equivalent to a barrier made from a low-$x$ Al$_x$Ga$_{1-x}$As alloy.
\bibitem{other-geometries} The samples shown here are
only a part of the full set of samples that we have investigated. We have
performed similar calculations for T-wires with asymmetric
parent QWs. We have also extended the numerical approach of
Ref.~\onlinecite{noi} to include the polarization charge
which forms at the interface between well and barrier materials due to
dielectric mismatch. All the investigated samples fall on the same curves
of Fig.~2.
\bibitem{foot2} The same value of the effective exciton Bohr
radius $a$ is reached in V- and T-wires for different values of the size
parameters: in general $L_V > L_T$. This is a consequence of the
single-particle confinement, which we find to be stronger in V- than in
T-wires when the barriers and the size parameters are taken to be 
the same ($L_V = L_T$).
\bibitem{foot4} For sample S2 of Ref.~\onlinecite{T2}, we calculate that
the shift between the lowest single-particle transitions of the T-wire
and the parent QW is underestimated by at least 8 meV.
Indeed, in Ref.~\onlinecite{T2} the electron and hole confinement energies
were estimated to be 20 and 1 meV respectively.
We find that the same quantities
must be larger than 25 and 4 meV.
\end{references}
\end{document}